\documentclass{JHEP3}
\usepackage{amssymb,amsfonts,amsmath,amsthm,latexsym,graphicx,verbatim,epsfig}
\newcommand{\be}{\begin{equation}}
\newcommand{\ee}{\end{equation}}
\newcommand{\bea}{\begin{eqnarray}}
\newcommand{\eea}{\end{eqnarray}}
\def\bse{\begin{subequations}}
\def\ese{\end{subequations}}
\newcommand{\IR}{\mathbb{R}} 

\def\IZ{\relax\ifmmode\hbox{Z\kern-.4em Z}\else{Z\kern-.4em Z}\fi}
\newcommand{\IS}{\mathbb{S}}

\newcommand{\non}{\nonumber \\}

\def\del{{\partial}}

\def\bi{\begin{itemize}} \def\ei{\end{itemize}}

\def\({\left(} \def\){\right)}
\def\[{\left[} \def\]{\right]}

\title{\center{Caged black hole thermodynamics: Charge, the extremal limit, and finite size effects}}

\author{
James~B.~Gilmore~and~Andreas~Ross
\\
Department~of~Physics,~Yale~University
\\
New~Haven,~CT~06520,~U.S.A.
\\
E-mail: \email{james.gilmore@yale.edu}, \
\email{andreas.ross@yale.edu} }

\author{
Michael~Smolkin
\\
Racah~Institute~of~Physics,~Hebrew~University
\\
Jerusalem~91904,~Israel
\\
E-mail: \email{smolkinm@phys.huji.ac.il}
}

\abstract{We extend the effective field theory treatment of the
thermodynamics of small compactified black holes to the case of
charged black holes. The relevant thermodynamic quantities are
computed to second order in the parameter $\lambda \sim
(r_0/L)^{d-3}$. We discuss how the addition of charge to a caged
black hole may delay the phase transition to a black string. In the
extremal limit, we construct an exact black hole solution which
serves as a check for our perturbative results. Finite size effects
are also included through higher order operators in the worldline
action. We calculate how the thermodynamic quantities are modified
in the presence of these operators, and show they enter beyond order
$\lambda^2$ as in the uncharged case. Finally, we use the exact
solution to constrain the Wilson coefficients of the finite size
operators in the extremal limit.}


\begin{document}

\section{Introduction}\label{sec:intro}

The physics of black objects in higher dimensional spacetimes can be
much richer than in four dimensions. In higher dimensions, it is
possible to find black objects of different horizon topologies and
transitions between these different phases. One extensively studied
example can be seen in the rotating stationary solutions of
\cite{Emparan:2001wn}, where both black hole (BH) and black ring
solutions exist. For particular combinations of mass and angular
momentum both phases can exist simultaneously and this has lead to a
rich phase diagram.

Another important system with black objects of different topologies
and interesting transitions is the BH-black string (BS) system; see
\cite{Kol:2004ww} and references therein. Here a spacetime with a
single compact dimension is considered, namely
$\IR^{d-2,1}\times\IS^1$, where $d$ is the total spacetime
dimension. Beyond $\IR^{d-1,1}$, this is the simplest possible
extension to the spacetime manifold, and thus presents a natural
candidate for study. In this paper we also calculate in this
spacetime. Note that in the case of a more elaborate manifold
compactification, one expects the transition physics to be similar
to the $\IS^1$ case.

In the BH-BS system, the BH and BS each possess distinct topologies
-- the BH horizon has a spherical topology $\IS^{d-2}$ and the BS
has a horizon topology $\IS^{d-3}\times \IS^1$. The stability of
both phases depends on the ratio of the two scales in the problem,
the mass $M$ and the size of the compact dimension $L$. For large
masses $GM \gg L^{d-3}$, a uniform BS is the only stable solution.
The analytical solution in this region of phase space is well-known
and exhibits a Gregory-Laflamme instability at lower masses
\cite{Gregory:1993vy}. For small masses $GM \ll L^{d-3}$ however,
the only stable phase is a BH. Since this BH is localized in a
compact dimension, it is referred to as a caged BH or Kaluza-Klein
(KK) BH. Moreover, in an intermediate mass region, there is an extra
phase corresponding to a non-uniform BS, where the non-uniformity
exists in the compact dimension. This phase may be stable or
unstable depending on the total number of spacetime dimensions.

Recently, the phase transitions between the uniform BS, the
non-uniform BS, and the caged BH have been discussed extensively. In
particular, numerical research of the BH/BS system has shown the BH
and the non-uniform BS phases merge at a topology changing
transition \cite{Wiseman:2002zc}. Although no closed form metric
exists for a caged BH, it is possible to calculate the properties of
\emph{small caged BHs} in a perturbation expansion in the parameter
$\lambda \sim (r_0/L)^{d-3}$. Here $r_0$ is the Schwarzschild radius
and $L$ is the size of the compact dimension. Note that in the
region where the BH/BS phase transition occurs, perturbative methods
break down and numerical methods or other non-perturbative
analytical approaches \cite{Kol:2002xz} must be employed.

There exist several methods to compute the perturbative expansion.
In \cite{Harmark:2003yz}, this problem was studied within adapted
coordinates in a single patch. In \cite{Karasik:2004ds}, the
thermodynamics were computed by applying the matched asymptotic
expansion method. Finally \cite{Chu:2006ce} used an effective field
theory (EFT) treatment to calculate the thermodynamics of neutral
caged BHs. This last approach appears to be the most systematic and
transparent and we employ it here. It applies the worldline EFT
methods used to describe extended objects in general relativity
(GR), originally proposed by Goldberger and Rothstein
\cite{Goldberger:2004jt, Goldberger:2006bd} (see
\cite{Goldberger:2007hy} for a pedagogical introduction).

In \cite{Chu:2006ce}, the EFT method was used to study the
thermodynamics of static caged BHs without charge up to $\mathcal
O(\lambda^2)$ in the perturbation expansion. The method was further
optimized in \cite{Kol:2007rx}, where a metric parametrization using
a temporal KK reduction was introduced. This parametrization
improves EFT calculations in static or weakly time dependent systems
\cite{Kol:2007bc}. The leading contributions to the thermodynamics
of small rotating caged BHs were also studied in \cite{Kol:2007rx}.

In addition to the thermodynamic calculations mentioned above, the
worldline EFT approach to GR has been successfully applied to the
post-Newtonian expansion of binary systems
\cite{Goldberger:2004jt,Kol:2007bc}. Spin degrees of freedom have
also been incorporated into the EFT framework
\cite{Porto:2005ac,Levi:2008nh} and gravitational radiation has been
studied \cite{Goldberger:2004jt,Cardoso:2008gn}. The self force on a
compact object was also computed \cite{Galley:2008ih} and absorptive
effects were introduced in \cite{Goldberger:2005cd,Porto:2007qi}.
Finally, the energy momentum tensor was considered in
\cite{Cannella:2008nr} and recently, \cite{Cannella:2009he} showed
how to constrain the three and four graviton vertices.

In this paper, we extend the perturbative studies of
\cite{Chu:2006ce} and \cite{Kol:2007rx} to electrically $U(1)$
charged caged BHs. Little is known about the phase diagram of this
system, but a richer phase structure can be expected, due to the
presence of the charge as an additional parameter. The stability of
the uniform charged BS has been considered \cite{Sarbach:2004rm} and
similar properties to the uncharged case were found. Recently,
\cite{Kleihaus:2009ff} used a Harrison transformation to study
charged black objects on KK spacetimes in Einstein-Maxwell-dilaton
gravity.

This work contains three main parts. First we compute all
thermodynamic quantities to $\mathcal O(\lambda^2)$ in the presence
of charge. Our results are valid for any value of the charge between
zero and its extremal value as long as $\lambda$ is small. Note that
to $\mathcal O(\lambda^2)$, the corrections stem purely from the
point particle description of the BH in the EFT framework. For
vanishing charge we recover the results of \cite{Chu:2006ce,
Kol:2007rx}.

Next, we construct an exact charged caged BH solution in the
extremal limit and compute its thermodynamics. In this limit, the
mass, tension, and electrostatic potential do not get renormalized,
i.e. they are simply given by their expressions in uncompactified
space. These results provide a further check of our perturbative
calculations.

Beyond $\mathcal O(\lambda^2)$, finite size effects are expected to
become important and we dedicate the last section of our paper to
them. In the EFT approach, finite size effects are simply
represented by higher order terms in the worldline action. We
evaluate the contributions of the leading order finite size
operators to various thermodynamic quantities for both uncharged and
charged caged BHs. The final results are expressed in terms of
undetermined matching coefficients, which need to be computed via a
matching procedure. Of particular note is the uncharged caged BH,
where there is a non-renormalization of the entropy at leading order
in the finite size operators. Finally, for a small charged caged BH
in the extremal limit, we undertake a matching calculation of the
finite size coefficients using our exact solution. Due to a
degeneracy in the thermodynamic relations, we are unable to uniquely
fix all three coefficients in the extremal limit.

\section{Effective field theory setup}\label{sec:EFTsetup}

In this section we will establish our conventions and the main
ingredients for our calculation. The general prescription of how to
set up an EFT for small KK BHs is explained in \cite{Chu:2006ce} and
\cite{Kol:2007rx} and we refer the reader to these papers for the
details such as the power counting.

We restrict our attention to the simplest compactification in higher
dimensional spacetime, that of $\IR^{d-2,1}\times\IS^1$. In this
spacetime a hierarchy of scales forms when we consider a small BH
with Schwarzschild radius $r_0\ll L$, where $L$ is the asymptotic
size of the compactified dimension. With such a hierarchy in place,
caged BHs can be treated within an EFT that allows their properties
to be calculated perturbatively in the expansion parameter
$\lambda$, which is defined by
\begin{equation}\label{eqn:lambda}
 \lambda :=\(\frac{r_0}{L}\)^{d-3}\zeta(d-3),
\end{equation}
with $\zeta(z)$ denoting the Riemann zeta function. The
Schwarzschild radius $r_0$ is given by \be
 r_0^{d-3}={16\pi G \, m_0 \over (d-2) \Omega_{d-2}} ~,~~~~
 \Omega_{d-2}={2\pi^{d-1 \over 2}\over \Gamma\left( d-1 \over 2
 \right)},
\ee where $m_0$ is the mass of the BH in an uncompactified
$d$-dimensional spacetime. In the language of quantum field theory
$m_0$ is called the bare mass of the BH. The charge of the BH is
$Q$. Coordinates on the $d$-dimensional spacetime are $x^\mu = (x^0,
\mathbf x)$ where $\mathbf x = (\mathbf x_\perp, z)$, with z
labeling the coordinate along $\IS^1$. We let Greek indices run over
all coordinates while Latin indices denote spatial components.

The setup of the EFT now proceeds as in \cite{Chu:2006ce,
Kol:2007rx} but we must also include the $U(1)$ gauge interaction.
Instead of obtaining a BH solution by solving the vacuum
Einstein-Maxwell equations, we integrate out the BH, i.e. we
integrate out the short distance scale $r_0$. This is achieved by
replacing the BH by an effective theory consisting of the worldline
of a massive charged particle coupled to the Einstein-Maxwell
system. The action is then given by $S =
S_{\textrm{EH-EM}}+S_{\textrm{BH}}$ with
 \bea
S_{\textrm{EH-EM}}[g_{\mu\nu},a_{\mu}]&=&\frac{-1}{16\pi
G}\int d^{d}x\sqrt{|g|}\left(R[g]+f^{\mu\nu}f_{\mu\nu}\right),
 \label{eqn:effactionEM}\\
S_{\textrm{BH}}[x, g_{\mu\nu}, a_{\mu}]&=&-m_0\int d\tau+Q\int
dx^{\mu}a_{\mu}+\ldots \ ,
 \label{eqn:BHaction}
 \eea
where $f_{\mu\nu}$ is the Maxwell field strength tensor, $a_{\mu}$
is the electromagnetic vector potential, and
$d\tau=\sqrt{g_{\mu\nu}dx^{\mu}dx^{\nu}}$. While the first two terms
in $S_{\textrm{BH}}$ describe a charged massive point particle, the
ellipsis denote higher order terms which encode finite size effects.
These terms are discussed in Section~\ref{sec:finitesize}.

Following \cite{Kol:2007rx} we exploit time translation symmetry and
apply the standard KK ansatz for the metric
 \begin{equation}\label{eqn:metric}
    ds^2=e^{2\phi}(dt-A_idx^i)^2-e^{-2\phi/(d-3)}\gamma_{ij} dx^idx^j,
 \end{equation}
reducing over the non-compact temporal coordinate. We can write
$\gamma_{ij} = \delta_{ij} + \sigma_{ij}$ so that flat Minkowski
spacetime is given by the ground state $\phi = A_i = \sigma_{ij} =
0$. This parametrization reorganizes the dynamical components of the
metric, i.e. $g_{\mu\nu}\rightarrow(\phi,A_{i},\sigma_{ij})$, into a
form which is advantageous from a computational perspective
\cite{Kol:2007rx,Kol:2007bc}. In addition, we also decompose the
electromagnetic $d$-vector potential into temporal and spatial
components $a_{\mu}=(\varphi,a_i)$. Note the distinction between the
notation for the KK gravitational scalar field $\phi$ and the
electromagnetic scalar field $\varphi$.

It is convenient to parameterize the worldline using coordinate time
$t = x^0$. Since our problem is static, we use the $d$-velocity
$dx^\mu/dt=(1,0)$ and we may fix the location of the BH at ${\mathbf
x} = 0$ without loss of generality. As a consequence, the fields
$A_i$, $a_i$, and $\sigma_{ij}$ cannot couple to the worldline as
long as we neglect finite size effects, and we will drop the
dependence on $x$ in $S_{\textrm{BH}}$. Staticity also implies there
is the additional symmetry of time reversal symmetry ${\mathbf T}$,
which means both vector fields $A_i$ and $a_i$, which are odd under
${\mathbf T}$, can be set to zero in any purely classical
calculation like the one here. In the static limit the action is
proportional to $\int dt$ and we will suppress this component from
this point. The resulting action takes the form
 \bea
 S_{\textrm{EH-EM}}[\phi, \sigma_{ij}, \varphi]&=&\frac{-1}{16\pi G}\int d^{d-1}\textbf{x} \sqrt{|\gamma|} \left(- R[\gamma] + {d-2\over
 d-3}(\partial\phi)^2-2e^{-2\phi}(\partial\varphi)^2\right),
 \label{eqn:KKEMaction}\\
 S_{\textrm{BH}}[\phi, \varphi]&=& -m_0 e^{\phi}+Q\varphi+\ldots\,,
 \label{eqn:KKBHaction}
 \eea
where $(\partial \phi)^2=(\del_i \phi) (\del_j \phi) \gamma^{ij}$
and similarly for $(\partial \varphi)^2$. We do not have to gauge
fix the action because the only propagators used in our calculations
are the $\phi$ and $\varphi$ propagators, and these can be easily
computed from (\ref{eqn:KKEMaction}).

The action we introduced above describes the dynamics at length
scales larger than $r_0$. In order to compute an observable measured
at infinity, we now have to integrate out the scale $L$. This is
performed by splitting up the fields into short and long wavelength
modes, e.g. we write $\phi = \bar \phi + \phi_L$ where $\bar \phi$
is the long wavelength mode and $\phi_L$ is the short wavelength
mode which we will integrate out. We will drop the index $L$ from
now on. The formal expression for the effective action is given by
\begin{equation}\label{eqn:integrateout}
 e^{i S_{\textrm{eff}}[\bar \phi, \bar \sigma_{ij}, \bar \varphi]} = \int \mathcal D \phi  \mathcal D \sigma_{ij} \mathcal D \varphi e^{i S[\bar \phi + \phi, \bar \sigma_{ij} + \sigma_{ij}, \bar \varphi + \varphi]},
\end{equation}
and clearly $S_{\textrm{eff}}$ will contain the standard
Einstein-Maxwell action $S_{\textrm{EH-EM}}[\bar \phi, \bar
\sigma_{ij}, \bar \varphi]$. The BH dynamics is contained in
$S_{\textrm{eff,BH}}$, which in practice is expanded in powers of
the long wavelength modes and the terms in this expansion are
computed via Feynman diagrams. These terms are then related to the
required observable. For example, the mass is simply given by the
constant term in $S_{\textrm{eff,BH}}$ without any couplings to the
long wavelength modes.

In \cite{Kol:2007rx} it was shown the KK parametrization of the
metric in (\ref{eqn:metric}) reduces the number of diagrams to be
calculated in the uncharged case. This remains true when we consider
charged compactified BHs, however there will be diagrams at
$\mathcal O(\lambda^2)$ which are not of an explicitly factorized
form. These diagrams arise due to a vertex with two electromagnetic
scalar fields $\varphi$ and one gravitational field $\phi$ derived
from (\ref{eqn:KKEMaction}).

\begin{figure}[tbp]
\centering
  \includegraphics{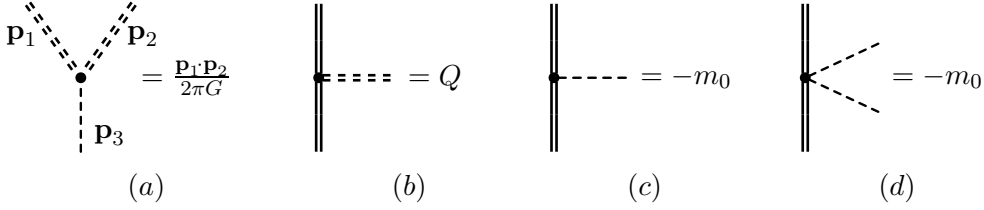}\\
  \caption{Feynman rules needed for the charged caged BH thermodynamics computations to $\mathcal{O}(\lambda^2)$. The single dashed lines correspond to the KK scalar field
  $\phi$,
  which is the propagating scalar mode in the gravitational sector and the double dashed lines denote the electromagnetic
  scalar $\varphi$. The Feynman rule (a) is given in momentum space, where we suppress
the delta function, and (b)-(d) are in coordinate
space.}\label{fig:FeynRule}
\end{figure}

We now proceed to write down the Feynman rules derived from the
action (\ref{eqn:KKEMaction}) and (\ref{eqn:KKBHaction}). The long
wavelength modes are not needed until Section~\ref{sec:finitesize}
so we set them to zero for now. The procedure for deriving the
Feynman rules is discussed in \cite{Chu:2006ce,Kol:2007rx} and we
simply summarize the results in Fig.~\ref{fig:FeynRule}. Note we
follow the conventions of \cite{Kol:2007rx} when deriving the
Feynman rules. In contrast to the gravitational couplings, it is
only possible to have one photon coupled to the worldline at a given
interaction point and we also have a $\phi \varphi \varphi$ vertex
as mentioned earlier. The propagator in position space for the
gravitational $\phi$ field is
 \be
 D_{\phi}(\mathbf{x}_\perp - \mathbf{x}'_\perp ;z-z') = \frac{8 \pi G}{L} \frac{d-3}{d-2}
 \sum^\infty_{n=-\infty} \int \frac{d^{d-2} \mathbf{k}_\perp}{(2\pi)^{d-2}}
 \frac{1}{\mathbf{k}_\perp^2 + (2\pi n/L)^2}
 e^{i  \mathbf{k}_\perp \cdot (\mathbf{x} - \mathbf{x}')_\perp +  2 \pi i n(z-z')/L},\label{eqn:KKscalarprop}
 \ee
and for the electromagnetic scalar $\varphi$ the propagator is
 \be
 D_{\varphi}(\mathbf{x}_\perp -\mathbf{x}'_\perp ;z-z')=-\frac{4\pi G}{L}
 \sum^\infty_{n=-\infty}\int\frac{d^{d-2} \mathbf{k}_\perp}{(2\pi)^{d-2}}
 \frac{1}{\mathbf{k}_\perp^2+(2\pi n/L)^2}
 e^{i\mathbf{k}_\perp \cdot (\mathbf{x} - \mathbf{x}')_\perp +  2 \pi i n(z-z')/L}.\label{eqn:EMscalarprop}
 \ee
In the Feynman diagrams, single dashed lines correspond to the
gravitational scalar field $\phi$, whereas double dashed lines
denote the electromagnetic scalar $\varphi$. The worldlines, which
are denoted by solid double lines, do not propagate so there are no
propagators associated with these lines.

\section{Thermodynamics of charged caged BHs from EFT} \label{sec:thermo}

To compute the thermodynamics of small charged caged BHs we first
calculate diagrammatically its ADM mass and the redshift factor. The
mass, for instance, is computed by integrating out short wavelength
modes of order $L$, as described in (\ref{eqn:integrateout}). Once
these quantities are at hand, one can calculate the remaining
thermodynamics.

\subsection{The ADM mass at $\mathcal O(\lambda^2)$} \label{sec:massadm}

\begin{figure}[tbp]
\centering
  \includegraphics{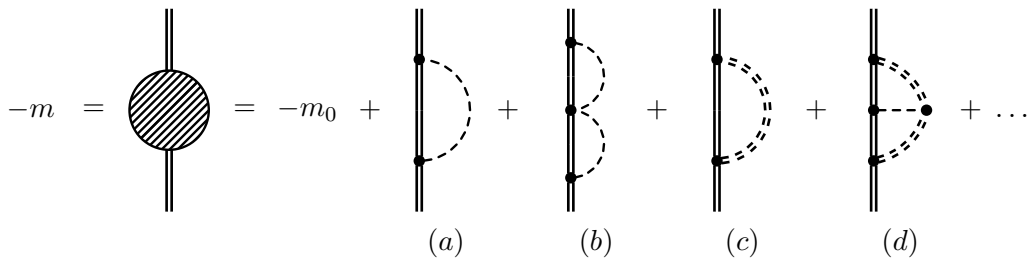}\\
  \caption{Diagrams that contribute to the ADM mass to $\mathcal O(\lambda^2)$.}\label{fig:Mrenorm}
\end{figure}

The ADM mass $m$ can be read off from the constant term in the
effective action $S_{\textrm{eff,BH}}$
\cite{Kol:2007rx} and to $\mathcal{O}(\lambda^2)$ this equals the
sum of classical vacuum diagrams shown in Fig.~\ref{fig:Mrenorm}. In
the uncharged case only diagrams in Fig.~\ref{fig:Mrenorm}$(a)$ and
$(b)$ contribute and these diagrams are calculated to be
\cite{Kol:2007rx}
\begin{eqnarray}
\textrm{Fig.
    \ref{fig:Mrenorm}$(a)$}&=&+{m_0 \over 2}\lambda\,,\\
\textrm{Fig.
    \ref{fig:Mrenorm}$(b)$}&=&-{m_0 \over 2}\lambda^2\,,
\end{eqnarray}
where the parameter $\lambda$ is given by (\ref{eqn:lambda}) and we
have used the Feynman rules of Section~\ref{sec:EFTsetup}, along
with (\ref{eqn:sumintiden}).

For the charged BH there are two additional diagrams to compute,
Fig.~\ref{fig:Mrenorm}$(c)$ and $(d)$. The computation of diagram
Fig.~\ref{fig:Mrenorm}$(c)$ is similar to
Fig.~\ref{fig:Mrenorm}$(a)$ and the result is
\begin{equation}\label{eqn:Mrenormfigc}
\textrm{Fig.
    \ref{fig:Mrenorm}$(c)$}=-\frac{Q^2}{4m_0}\frac{d-2}{d-3}\,\lambda.
\end{equation}
It can be understood in terms of the total electrostatic energy
$E={1 \over 2}\sum_{i} Q \, \varphi_i(O)$, where the sum runs over
all images of the BH in the covering space and $\varphi_i(O)$
represents the electrostatic potential created by the image $i$ at
the location of the BH.

Diagram Fig.~\ref{fig:Mrenorm}$(d)$ is slightly more complicated, so
we calculate it here explicitly. The symmetry factor we divide by is
2. We can now write the amplitude
\begin{equation}\label{eqn:Mrenormfigda}
  -(4 \pi G Q)^2\,\frac{d-3}{d-2}\,m_0\,\,\, \int\!\!\!\!\!\!\!\!\!\!\!\!\sum_{\,\,\textbf{p}_1,\textbf{p}_2}\frac{2 \, \textbf{p}_1\cdot
   \textbf{p}_2}{\textbf{p}_1^2 \, \textbf{p}_2^2(\textbf{p}_1+\textbf{p}_2)^2},
\end{equation}
where ${\mathbf p} = \left(\textbf{p}_{\perp}, \frac{2 \pi
n}{L}\right)$ and
$$\int\!\!\!\!\!\!\!\!\!\! \hspace*{0.6pt} \sum_{\textbf{p}\,\,\,\,\,\,} = \frac{1}{L} \sum_{n=-\infty}^{\infty}\int\frac{d^{d-2}\textbf{p}_{\perp}}{(2\pi)^{d-2}}.$$
By using the identity $2 \, \textbf{p}_1\cdot \textbf{p}_2=
(\textbf{p}_1+\textbf{p}_2)^2-\textbf{p}_1^2-\textbf{p}_2^2$ and
then subsequently redefining the momentum, we find this amplitude
can be written in a factorized form
\begin{equation}\label{eqn:Mrenormfigdb}
   +(8 \pi G Q)^2\,\frac{d-3}{d-2}\,m_0 \left[\frac{1}{2 L}\sum_{n=-\infty}^{\infty}\int\frac{d^{d-2}\textbf{p}_{\perp}}{(2\pi)^{d-2}}
   \frac{1}{\textbf{p}_{\perp}^2+(2\pi n/L)}\right]^2,
\end{equation}
and with the identity (\ref{eqn:sumintiden}) it leads to
\begin{equation}\label{eqn:Mrenormfigd}
      \textrm{Fig.
    \ref{fig:Mrenorm}$(d)$}=+\frac{Q^2}{4m_0}\frac{d-2}{d-3}\,\lambda^2\, .
\end{equation}

Combining these results gives the following expression for the ADM
mass to second order in $\lambda$
\begin{equation}\label{eqn:mass}
    m=m_0\left[1-\frac{1}{2}\,\eta^2\,\lambda+\frac{1}{2}\,\eta^2\,\lambda^2+\ldots\right].
\end{equation}
The variable $\eta$ in the above expression plays the role of the
order parameter and is given by
\begin{equation}\label{eqn:eta}
\eta=\sqrt{1-\frac{1}{2}\frac{d-2}{d-3}\frac{Q^2}{m_0^2}}.
\end{equation}
It vanishes in the extremal limit when
\begin{equation} \label{eqn:extcondition}
 \frac{1}{2}\frac{d-2}{d-3}\,Q^2=m_0^2
\end{equation}
holds and reaches its maximum value, $\eta=1$, for an uncharged BH.
Note also, corrections to the mass of an extremally charged BH
vanish to second order identically, namely the ADM mass is left
unrenormalized to $\mathcal{O}(\lambda^2)$ in the extremal limit. As
we show in Section~\ref{sec:exact_extremalsol}, it turns out such a
non-renormalization of the ADM mass must hold to any order in the
perturbation expansion.

\subsection{Thermodynamics} \label{subsec:Thermo}

In the previous subsection we computed the BH mass (\ref{eqn:mass})
in terms of the charge $Q$, the compactification parameter $L$, and
the bare mass $m_0$, however this relation does not fully determine
the thermodynamics of the system. In this section we will compute
the remaining thermodynamic quantities, the temperature $T$, the
entropy $S$, the electrostatic potential $\Phi$, and the tension
$\hat \tau$ \cite{Traschen:2001pb}. We also give the expression for the Gibbs potential
$G$.

\begin{figure}[tbp]
\centering
  \includegraphics{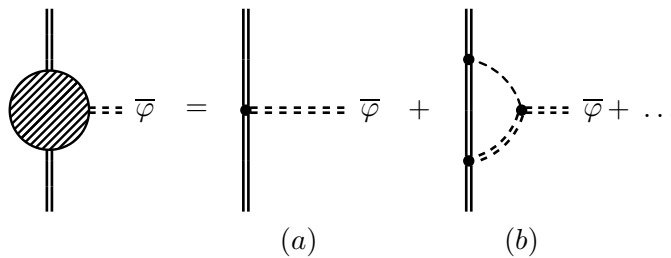}\\
  \caption{Diagrams that contribute to the asymptotic value of the electric charge.}\label{fig:Qrenorm}
\end{figure}

First, we will clarify the notion of the charge $Q$. In our
calculations there is no distinction between the bare and asymptotic
values of the electric charge $Q$. Intuitively such a
non-renormalization stems from the electric flux conservation
requirement. Technically, the asymptotic value of the electric
charge can be read off the following linear term in the effective
action
\begin{equation}
 S_{\textrm{eff,BH}}[\bar \phi, \bar \sigma_{ij}, \bar \varphi]\supset + Q \bar \varphi \, .
\end{equation}
This term is the sum of Feynman diagrams like those of
Fig.~\ref{fig:Qrenorm} with $\bar \varphi$ set to a constant, in
which case the momentum flowing into the diagram vanishes. All
couplings involving gravitons and photons are derived from the
second term in (\ref{eqn:effactionEM}). Since all of these vertices
contain derivatives acting on the electromagnetic scalar field $\bar
\varphi$, they all vanish in the zero momentum limit. The only
possible vertex for a photon coupling to the worldline is
Fig.~\ref{fig:FeynRule}(b) which cannot include any additional
gravitons.  Thus only the leading order diagram is nonzero and the
sub-leading diagrams such as Fig.~\ref{fig:Qrenorm}(b) vanish. This
means the asymptotic value of the electric charge is equal to its
bare value $Q$. The inclusion of finite size effects does not alter
this conclusion, since all higher order operators with
electromagnetic interactions include derivatives acting on $\bar
\varphi$.

\begin{figure}[tbp]
\centering
  \includegraphics{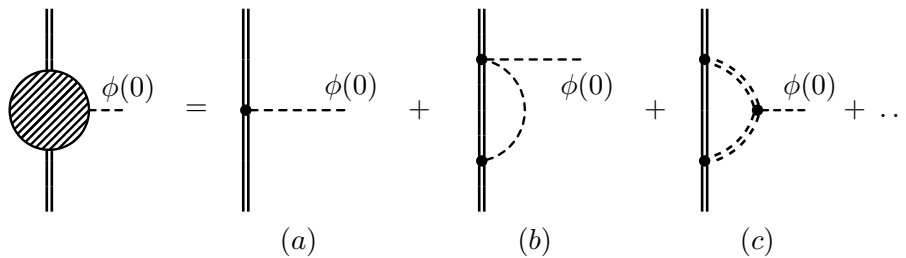}\\
  \caption{Diagrams required to compute the redshift factor $R$ to $\mathcal{O}(\lambda^2)$.
  Here, $\phi(0)$ denotes that the $\phi$ field propagates back to $\textbf{x}=0$ on the BH worldline
  but does not have a vertex.}\label{fig:Phi}
\end{figure}

Next we compute the asymptotic temperature. It is canonically
conjugate to time and hence transforms inversely to $t$, namely
$T/T_0=t_0/t$. Therefore one can relate the asymptotic temperature
$T$ to the local temperature $T_0$ using the redshift,
 \be
 T=R\,T_0\, .
 \label{eqn:tempeqn}
 \ee
As long as we do not include finite size effects, the horizon will
not be deformed and the local temperature will coincide with the
temperature of an uncompactified $d$-dimensional
Reissner-Nordstr\"{o}m BH \cite{Myers:1986un},
 \be \label{eq:T0rn}
 T_0(m_0,Q)=T(L \rightarrow \infty) = \frac{d-3}{d-2}\,\frac{m_0}{S_0}\,\eta\,,
 \ee
where $\eta$ is the order parameter defined in (\ref{eqn:eta}) and
$S_0$ is the entropy of an uncompactified $d$-dimensional charged BH
given by
 \begin{equation}
  S_0(m_0,Q)=\frac{k_B}{4G}\,\Omega_{d-2}\,r_{+}^{d-2} \, ,~~~~~~
  r_{+}^{d-3}=\frac{8\pi G m_0}{(d-2)\Omega_{d-2}}(1+\eta) \, .
  \label{eqn:S0}
 \end{equation}
Note the local temperature $T_0$ in (\ref{eq:T0rn}) and the
expression for $S_0$ in (\ref{eqn:S0}) are correct away from the
extremal limit where subtleties might arise in the definition of
temperature and entropy \cite{Hawking:1994ii}. The redshift factor
$R$ in (\ref{eqn:tempeqn}) is given by
 \be
 R ={t_0 \over t}=\sqrt{g_{00}(0)}=e^{\phi(0)} \, ,
 \label{eqn:redshiftformula}
 \ee
where $\phi(0)$ is defined in terms of the Feynman diagrams given in
Fig.~\ref{fig:Phi}. We now calculate these diagrams to obtain $R$.

The first two diagrams, Fig.~\ref{fig:Phi}$(a)$ and $(b)$ are charge
independent and their value is given by \cite{Kol:2007rx}
 \bea
  \textrm{Fig.~\ref{fig:Phi}$(a)$}&=&-\lambda,\\
  \textrm{Fig.~\ref{fig:Phi}$(b)$}&=&+\lambda^2.
 \eea
Evidently, the computation of the redshift factor is only modified
at $\mathcal{O}(\lambda^2)$ by diagram Fig.~\ref{fig:Phi}$(c)$. The
steps to compute this additional diagram are similar to those used
to compute Fig.~\ref{fig:Mrenorm}$(d)$. It is merely obtained by
dividing the result of Fig.~\ref{fig:Mrenorm}$(d)$ by $-m_0$. This
in turn yields
 \begin{equation}\label{eqn:phic}
  \textrm{Fig.~\ref{fig:Phi}$(c)$}=-\frac{Q^2}{4m_0^2}\frac{d-2}{d-3}\lambda^2.
 \end{equation}
Combining the above results, we obtain the redshift factor
\begin{equation} \label{eq:redshift}
 R = 1-\lambda+\left(1+\frac{\eta^2}{2}\right)\lambda^2+\ldots \, .
\end{equation}
In the extremal limit $\eta = 0$, the redshift factor agrees with
the exact solution constructed in
Section~\ref{sec:exact_extremalsol}. For the temperature to
$\mathcal{O}(\lambda^2)$ in the case of a charged compactified BH
the result is then
\begin{equation}
    \label{eqn:temperature}
    T=T_0\left[1-\lambda+\left(1+\frac{\eta^2}{2}\right)\lambda^2+\ldots\right].
\end{equation}

One can close the thermodynamic equations by considering the entropy
of the charged caged BH. Since the contribution of finite size
operators are always beyond $\mathcal{O}(\lambda^2)$ in the
perturbative expansion, we can deduce horizon deformations due to
the presence of the compact dimension can be neglected. In other
words, the BH horizon is unaltered by the compactified dimension to
$\mathcal{O}(\lambda^2)$ and its area is fixed by the values of
$m_0$ and $Q$ only. Since the entropy of the BH is proportional to
the area of the horizon, we can write down the entropy of the
charged caged BH to $\mathcal{O}(\lambda^2)$ as follows
\begin{equation}
 S=S_0(1+0\,\lambda+0\,\lambda^2+\ldots),
\end{equation}
where $S_0$ is the entropy of an uncompactified $d$-dimensional
charged BH given by (\ref{eqn:S0}). In Appendix
\ref{Sec:Helmholtzappendix} we present an independent derivation of
the thermodynamic relations using the Helmholtz free energy, which
does not use the uncompactified entropy as an input.

From the first law of thermodynamics
\begin{equation}\label{eqn:1stlaw}
 dm=TdS+\Phi dQ + \hat\tau dL,
\end{equation}
one can derive the expression for the electrostatic potential
 \bea
 \Phi&=&\( {\partial m \over \partial Q}\)_{S,\,L}
 = {\partial(m,S,L) \over \partial (Q,m_0,L)}\left({\partial(Q,S,L) \over \partial(Q,m_0,L)}\right)^{-1}\nonumber \\
 &=& \( {\partial m \over \partial Q}\) -\( {\partial m \over \partial m_0}\)
 \( {\partial S \over \partial Q}\) \( {\partial S \over \partial m_0}\)^{-1}
 \, ,
 \eea
where we have used the standard change of thermodynamic variables in
terms of Jacobians \cite{LL}. Substituting explicit expressions for
$S$ and $m$ from (\ref{eqn:mass}) and (\ref{eqn:S0}) respectively,
we obtain
\begin{equation}\label{eqn:Phi}
     \Phi=\Phi_0\left[1+\eta\,\lambda-{\eta \over
     2}\,(2-\eta)\lambda^2+\ldots\right],
\end{equation}
where $\Phi_0=(m_0/Q)(1-\eta)$.

To compute the tension $\hat{\tau}$, we use the Smarr relation
\begin{equation} \label{eq:smarr}
 (d-3)(m-\Phi Q)=(d-2)TS+\hat\tau L,
\end{equation}
which gives
\begin{equation}\label{eqn:tensionfinal}
    {\hat\tau L \over m_0}={1 \over
    2}(d-3)\eta^2\lambda-(d-3)\eta^2\lambda^2 +\ldots \, .
\end{equation}
We have explicitly checked this result for the tension via a direct
diagrammatic computation with an external $\bar \sigma_{zz}$, see
Section~\ref{sec:finitesize} for further details on the procedure.

The Gibbs free energy
\begin{equation}
G=G(T,\Phi,L)=m-TS-\Phi Q,
\end{equation}
which implies $dG=-SdT-Qd\Phi+\hat\tau dL$. Although we did not need
to use the Gibbs free energy above, all the thermodynamics can be
derived from $G$ by forming the appropriate differentials. The
result reads
\begin{equation}\label{eqn:GibbsPotentia}
    G(m_0,Q,L)=G_0\left[1+\left({d-2 \over 2}\,\eta - 1\right)\lambda+\left( 1-(d-2)\eta+{\eta^2 \over 2} \right)\lambda^2 +\ldots \right],
\end{equation}
where $G_0=\eta\,m_0/(d-2)$ is the Gibbs potential of a charged BH
in the uncompactified spacetime $\IR^{d-1,1}$.

Examining our thermodynamic calculations, we recover the results for
an uncharged caged BH of \cite{Chu:2006ce,Kol:2007rx} in the
vanishing charge limit $\eta = 1$. On the other hand, when the
extremality condition $\eta = 0$ holds, we have found to $\mathcal
O(\lambda^2)$ the mass and the electrostatic potential remain
unrenormalized and the redshift $R$ is given by (\ref{eq:redshift})
with $\eta = 0$. In addition, according to (\ref{eqn:tensionfinal}),
the tension of an extremally charged BH vanishes to second order
identically. Physically, this result states that electrostatic
repulsion exactly cancels gravitational attraction to this order in
$\lambda$. These results are in agreement with the exact solution we
construct for an extremally caged BH in the next section.

\section{Charged caged BHs in the extremal limit} \label{sec:exact_extremalsol}

It has been known for some time that Einstein-Maxwell theory admits
static multi-BH solutions \cite{Papapetrou:1947ib}. The existence of
these solutions can be attributed to the balance of electromagnetic
and gravitational forces when the extremal condition
(\ref{eqn:extcondition}) is satisfied. In this section we follow
\cite{Gibbons:1976ue} and \cite{Myers:1986rx} and build the
thermodynamics for our extremally charged caged BH. This will allow
us to verify the perturbative expansions of the previous section in
the extremal limit.

\subsection{Exact solution}

The action we employ is the Euclidean Einstein-Maxwell
action\footnote{In this section we change our notation to emphasize
that we are not working in the EFT but in full GR, namely all scales
are present at all times. Also, since our problem is static,
switching between Lorenzian and Euclidean signatures presents no
difficulties and is given by $t \rightarrow it$.} supplemented with
the Gibbons-Hawking boundary term \cite{Gibbons:1976ue}. This
additional term is included to obtain an action depending only on
the first derivatives of the metric. We therefore have
 \begin{equation} \label{eqn:HE+GH_action}
    S_E^{full}[g_{\mu\nu}^f,\mathcal{A}_{\mu}]=\frac{-1}{16\pi G}\int d^{d}x\sqrt{g_f}\left(\mathcal{R}[g_{\mu\nu}^f]
    -\mathcal{F^{\mu\nu}F_{\mu\nu}}\right)
    -\frac{1}{8\pi G}\int_{ \partial \mathcal{M}} \sqrt{h}\, (\mathcal{K}-\mathcal{K}_0) ,
 \end{equation}
where $g_{\mu\nu}^f$ and $\mathcal{A}_{\mu}$ are the full metric and
the full electromagnetic $d$-vector potential including all length
scales involved in the problem; $\mathcal{R}$ is the Ricci scalar
and
$\mathcal{F}_{\mu\nu}=\partial_{\mu}\mathcal{A}_{\nu}-\partial_{\nu}\mathcal{A}_{\mu}$
is the field strength tensor, $h_{ij}$ represents an induced metric
on $\partial\mathcal{M}$, and $\mathcal{K}$ and $\mathcal{K}_0$ are
the trace of the second fundamental form of the boundary for a
particular metric and a flat metric respectively.

\begin{figure}[tbp]
\centering
  \includegraphics{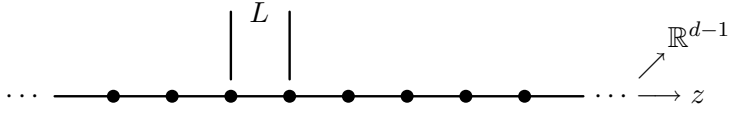}\\
  \caption{Schematic of exact solution construction.}\label{fig:exactsol}
\end{figure}

For the caged BH setup, the gravitational field can be thought to
originate from the BH and its associated images in the covering
space. In other words, the setup is equivalent to the case in which
an infinite number of identical equidistant BHs lie on a common axis
(the $z$-axis) of a $d$-dimensional uncompactified spacetime, see
Fig.~\ref{fig:exactsol}. This arrangement has the following solution
\cite{Myers:1986rx}
 \be
 \label{eqn:ExactMetric}
 ds^2=g_{\mu\nu}^f dx^{\mu}dx^{\nu}
 =H^{-2}d\tau^2+H^{2/(d-3)}\left( d\rho^2+\rho^2 d\theta^2 + \rho^2 \sin^2 \theta \, d\Omega_{d-3}^2\right),
 \ee
where $H$ extremizes the action (\ref{eqn:HE+GH_action}) and is
given below, $d\Omega_{d-3}^2$ denotes the metric on the unit
$(d-3)$-sphere and $\tau$ corresponds to the Euclidean time. The
Maxwell field strength is $\mathcal{F}=d\mathcal{A}$, where
 \be
 \label{eqn:1form}
 \mathcal{A}=-\sqrt{{d-2\over 2(d-3)}}\textrm{sign}(Q) \, H^{-1}d\tau .
 \ee
The function $H$ is harmonic on $\IR^{d-1}$, with poles
corresponding to the locations of the event horizons of the BHs and
is given by
 \be
 H=1+{ r_0^{d-3} \over 2} \sum^{\infty}_{n=-\infty} {1 \over (\mathbf{x}_\perp^2+(z - n L)^2)^{(d-3)/2}}\,,
 \label{eqn:H}
 \ee
where $|\mathbf{x}_\perp|=\rho\sin\theta$ and $z=\rho\cos\theta$.

A comment should be made regarding the gauge choice of the 1-form
$\mathcal{A}$. As can be seen from (\ref{eqn:1form}),
$\mathcal{A}_0=0$ at any event horizon. Since $d\tau$ is not defined
on the horizon, this choice of gauge avoids the possibility of a
singularity in the 1-form at any event horizon. The potential at
infinity does not vanish in this gauge, so the electrostatic
potential is given by
 \be\label{eqn:extremePhi}
 \Phi=\mathcal{A}_0(\rho=0)-\mathcal{A}_0(\infty)=\sqrt{{d-2\over
 2(d-3)}}\textrm{sign}(Q)\,,
 \ee
where we have chosen to evaluate $\Phi$ at the horizon located at
the origin.

If $L\rightarrow\infty$ in (\ref{eqn:H}), so that only the $n=0$
term remains, we recover the standard Reissner-Nordstr\"{o}m
extremal solution written in the isotropic radial coordinate $\rho$.
Its relation to the Schwarzschild coordinate $r$ is given by
\begin{equation}
    \rho=\left(r^{d-3}-{1 \over 2}\, r_0^{d-3}\right)^{1/(d-3)}\,.
\end{equation}

\subsection {Thermodynamics}

In what follows we evaluate the Euclidean action
(\ref{eqn:HE+GH_action}) on the solutions (\ref{eqn:ExactMetric})
and (\ref{eqn:1form}). We find the Euclidean action vanishes and
this leads to the conclusion that the Gibbs potential of an extremal
charged caged BH equals zero, via the relation
 \be
 -\beta G(m_0,L) = \ln Z = -S_E^{full}[g_{\mu\nu}^f,\mathcal{A}_{\mu}],
 \ee
where $\beta$ is the inverse Euclidean temperature and $Z$ is the
partition function. As shown in \cite{Hawking:1994ii}, due to the
topology change of the spacetime manifold, the Euclidean extremal
Reissner-Nordstr\"{o}m solution has two boundaries: at infinity (as
in the non-extremal case) and at the horizon. We now evaluate the
contribution these boundaries make to the Euclidean action.

We first consider the boundary at infinity and we choose
$r_{\perp}=const$, where $r_{\perp}=|\mathbf{x}_\perp|$, to
represent this surface. This boundary is compact and has the
topology $\IS^1 \times \IS^1 \times \IS^{\,d-3}$. The series in
(\ref{eqn:H}) can be summed explicitly if $d$ is odd, that is
$d=2j+5, j=0,1,2,\ldots\,$. In this case
 \bea
 H&=&1+{r_0^{d-3} \over 2}{(-1)^j \over j!} {d^j \over d(\mathbf{x}_\perp^2)^j}
 \sum^{\infty}_{n=-\infty} {1 \over \mathbf{x}_\perp^2+(z - n L)^2}
 \nonumber \\
 &=&1+r_0^{d-3}{ \pi  \over 2 L}{(-1)^j \over j!}{d^j \over
d(\mathbf{x}_\perp^2)^j}\left[
 {1 \over |\mathbf{x}_\perp|} {\sinh \left( {2\pi |\mathbf{x}_\perp| \over L} \right)
 \over \cosh\left( {2\pi |\mathbf{x}_\perp| \over L} \right)- \cos \left( {2\pi z \over L} \right) }
 \right]\nonumber \\
 &=& 1+{d-3 \over 2(d-4)}~ {\Omega_{d-2} \over \Omega_{d-3}} ~ {r_0 \over L}
 \left( r_0 \over |\mathbf{x}_\perp| \right)^{d-4},
 \eea
where in the last equality we suppressed exponentially small terms
of order $\sim\exp(-L/|\mathbf{x}_\perp|)$. These terms do not
contribute further in our computations. Further, since the leading
order term contains no information about parity, one can show this
result is valid for any $d$. Therefore we have
 \bea
 ds^2&=&\left( 1+{d-3 \over 2(d-4)}~ {\Omega_{d-2} \over \Omega_{d-3}} ~
 {r_0 \over L} \left( r_0 \over r_{\perp} \right)^{d-4}\right)^{-2}d\tau^2
 \nonumber \\
 &+&\left( 1+{d-3 \over 2(d-4)}~ {\Omega_{d-2} \over \Omega_{d-3}} ~ {r_0 \over L}\left( r_0 \over r_{\perp} \right)^{d-4}\right)^{2/(d-3)}
 \left( dz^2 + dr^2_{\perp} + r^2_{\perp} \,d\Omega_{d-3}^2\right),
 \nonumber \\
 \mathcal{F}&=&d\mathcal{A}=-\sqrt{{(d-2)(d-3)\over 8}} \, {1 \over  L}~ {\Omega_{d-2} \over \Omega_{d-3}} ~
 \left(r_0\over r_{\perp}\right)^{d-3}H^{-2}~dr_{\perp}\wedge dt.
 \label{eqn:extremesol}
 \eea

One can evaluate the corresponding Gibbons-Hawking boundary term
relying on the identity
 \be
 \int_{ \partial \mathcal{M}} \sqrt{h} \mathcal{K} = {\partial \over \partial n} \int_{ \partial
 \mathcal{M}}\sqrt{h},
 \ee
where $(\partial / \partial n) \int_{ \partial \mathcal{M}}
\sqrt{h}$ is the derivative of the area of $\partial \mathcal{M}$ as
each point of the boundary is moved an equal distance along the
outward unit normal $n$. Thus in our case we get
 \bea
 \int_{ r_{\perp}=const} \sqrt{h}\,\mathcal{K} &=& H^{-1/(d-3)}{d \over
 dr_{\perp}} \left( \beta L  \, \Omega_{d-3}\,H^{1/(d-3)}r_{\perp}^{d-3}\right)
 \nonumber \\
 &=& \beta L  \, \Omega_{d-3}\left[(d-3)r_{\perp}^{d-4}-{\Omega_{d-2} \over \Omega_{d-3}} ~
 {r_0^{d-3} \over 2\,L}\right].
 \eea
For flat space $\mathcal{K}_0=(d-3)H^{-1/(d-3)}r_{\perp}^{-1}$,
therefore
 \be
 \int_{ r_{\perp}=const} \sqrt{h}\,\mathcal{K}_0=\beta L  \,
 \Omega_{d-3}(d-3)r_{\perp}^{d-4}.
 \ee
Combining these relations together yields
 \be
 {1 \over 8\pi G}\int_{ r_{\perp}=const}\sqrt{h}\,
 \left(\mathcal{K}-\mathcal{K}_0\right)
 =- {\beta m_0 \over d-2}.
 \label{eqn:InfBoundary}
 \ee

We now turn to the boundary at the horizon, given by $\rho=0$. To
compute this boundary contribution, we consider the hypersurface
$\rho=const$ and then take the limit $\rho\rightarrow 0$ at the end
of the calculation. An outward-pointing normal vector for such a
hypersurface is $n_{\mu}=-\sqrt{g_{\rho\rho}}~\delta_{\mu\rho}$ (no
sum on $\rho$) and the induced metric $h_{\mu\nu}$ is given by
 \be
 ds^2=h_{\mu\nu} dx^{\mu}dx^{\nu}
 =H^{-2}d\tau^2+H^{2/(d-3)}\rho^2\left( d\theta^2 + \sin^2 \theta \, d\Omega_{d-3}^2\right),
 \ee
and the covariant derivative of $n_{\mu}$ is
 \be n_{\nu;\,\mu}=-\sqrt{g_{\rho\rho}}~
 \left(\delta_{\mu\rho}\,\delta_{\nu\rho}\,\partial_{\rho}\ln\sqrt{g_{\rho\rho}}
 + \delta_{\mu\theta}\,\delta_{\nu\rho}\,\partial_{\theta}\ln\sqrt{g_{\rho\rho}}
 -\Gamma_{\mu\nu}^{\rho}\right).\ee
We can now evaluate the trace of the second fundamental form on the
boundary
 \be
 \mathcal{K}=h^{\nu\mu}\,n_{\nu;\,\mu}=-H^{-1 / (d-3)}\left( {d-2 \over \rho} + {1 \over d-3}\, {\del_{\rho} H \over H}
 \right).
 \ee
From the definition (\ref{eqn:H}) of $H$, we get the following
expansion in the vicinity of the horizon
 \bea
 H&=& {1 \over 2} \left( r_0 \over \rho \right)^{d-3}
 \left(1+\mathcal{O}(\rho^{d-3})\right) \, ,
 \nonumber \\
 {\del_{\rho} H \over H}&=&
 { 3-d \over \rho}\left(1+\mathcal{O}(\rho^{d-2})\right).
 \eea
The regulator in this case is given by $\mathcal{K}_0=-(d-2)H^{-1 /
(d-3)}/\rho$ and we therefore obtain
 \be
 {1 \over 8\pi G}\int_{ \rho=const}\sqrt{h}\,
 \left(\mathcal{K}-\mathcal{K}_0\right)
 = {\beta\,\Omega_{d-2} \over 8\pi G  }\,\rho^{d-3}
 \left(1+\mathcal{O}(\rho^{d-2})\right) \, .
 \ee
As a result, the horizon boundary term vanishes when $\rho$ tends to
zero, that is
 \be
 {1 \over 8\pi G}\int_{ \rho=0}\sqrt{h}\,
 \left(\mathcal{K}-\mathcal{K}_0\right)=0 \, .
 \label{eqn:HorBoundary}
 \ee

In order to complete the calculation of the Euclidean action, one
has to compute the Einstein-Maxwell terms in
(\ref{eqn:HE+GH_action}). Since the scalar curvature $\mathcal{R}$
can be evaluated from the Einstein field equation
 \be
 \mathcal{R}_{\mu\nu}-{1\over 2}g_{\mu\nu}\mathcal{R}=8\pi G\,\mathcal{T}_{\mu\nu} \Rightarrow
 ~ \mathcal{R}={ d-4 \over d-2}\,\mathcal{F^{\mu\nu}F_{\mu\nu}},
 \label{eqn:scalar_curvature}
 \ee
where
 \be
 \mathcal{T}_{\mu\nu}={1 \over 4\pi G}(\mathcal{F_{\mu}^{~\alpha}F_{\nu \alpha}}
 -{1\over 4}g_{\mu\nu}\mathcal{F^{\alpha\beta}F_{\alpha\beta}}),
 \ee
is the energy-momentum tensor of the Maxwell field, it is enough to
evaluate the electromagnetic part of the action.

For this purpose, recall that for a solution of the Maxwell
equations $\mathcal{F}^{\mu\nu}_{~~~;\,\nu}=0$, one can rewrite the
Maxwell field Lagrangian density as
$\mathcal{F_{\mu\nu}F^{\mu\nu}}=(2\mathcal{F^{\mu\nu}A_{\mu}})_{;\,\nu}$.
Combining with (\ref{eqn:extremesol}) gives
 \be
 \frac{1}{16\pi G} \int_{\mathcal{M}} \sqrt{g}
  \mathcal{F_{\mu\nu}F^{\mu\nu}}={1 \over 8 \pi G} \int \mathcal{F^{\mu\nu}A_{\mu}}d\Sigma_{\nu}= -{ \beta m_0 \over
  2}.
 \label{EM_action}
 \ee

We are now in a position to compute the Gibbs free energy.
Substituting (\ref{eqn:InfBoundary}), (\ref{eqn:HorBoundary}),
(\ref{eqn:scalar_curvature}), and (\ref{EM_action}) into
(\ref{eqn:HE+GH_action}) yields
 \be
 -\beta G(m_0,L) = -S_E^{full}=0.
 \label{Gibbs}
 \ee
Since $G(m_0,L)=m-TS-\Phi Q$ this implies
 \be
 m-\Phi Q=TS,
 \label{fullmassrelation}
 \ee
where $m$ is the total energy (mass) of the system. However, by the
generalized Smarr's relation $(d-3)(m-\Phi Q)=(d-2)TS+\hat{\tau} L$.
Thus $-TS=\hat{\tau} L$. Since all the quantities involved in the
last equality are positive definite and $L$ does not vanish, we
conclude in the extremal case
 \be
 T S = \hat{\tau} = 0,
 \label{eqn:temp_tension}
 \ee
so the tension of the system is zero. Substituting this result back
into (\ref{fullmassrelation}) and using the extremality condition
(\ref{eqn:extcondition}), we see the mass of the system is
unrenormalized
 \be
 m=\Phi Q=m_0.
 \label{eqn:fullmass}
 \ee

There remains one last thermodynamic quantity of interest. The
redshift factor (\ref{eqn:redshiftformula}), is easily computed from
our exact extremal solution and we have
 \be
 \sqrt{g_{00}^f(\mathbf{x})}=H^{-1}(\mathbf{x})~\Rightarrow~
 R=\sqrt{g_{00}(O)}=(1+\lambda)^{-1} \, ,
 \label{eqn:ExactRedshiftextreme}
 \ee
where in the last equality we excluded the $n=0$ term in the
definition of $H$, since it corresponds to the short wavelength
scale of order $r_0$. This term is a pure infinity and is also
present in the EFT approach, but in that case it is set to zero via
dimensional regularization. For our purposes here, it can be ignored
and thus does not contribute to the value of $g_{00}$. Expanding the
last equation in $\lambda$, one recovers (\ref{eq:redshift}) for
$\eta=0$.

The results in (\ref{eqn:temp_tension}), (\ref{eqn:fullmass}), and
(\ref{eqn:ExactRedshiftextreme}) summarize the major outcomes of
this section. In particular, these results verify the
non-renormalization statements obtained perturbatively in
Section~\ref{sec:thermo} via the EFT approach.

\section{Finite size effects}\label{sec:finitesize}

In this section we add non-minimal couplings to account for finite
size effects and study the resulting thermodynamics. We start with
the uncharged case and then proceed to charged caged BHs. We
determine the order the non-minimal operators enter the perturbative
expansion for charged BHs. The existence of the exact solution in
the extremal case allows us to undertake a matching calculation for
the Wilson coefficients of the finite size operators in this limit.

\subsection{Uncharged caged BHs}

To incorporate the finite size operators into the action, we need to
add all possible non-minimal operators which respect diffeomorphism
and reparametrization invariance. These operators must use proper
time throughout and can involve combinations of the geometric
invariants, i.e. the Ricci scalar $R$, the Ricci tensor $R_{\mu
\nu}$, and the Riemann tensor $R_{\mu \nu \rho \sigma}$. Higher
derivative combinations of these terms are also possible.

As usual in an EFT, the derivative expansion is truncated at a given
order determined by the accuracy required in the calculation. In the
uncharged case \cite{Goldberger:2004jt} it was shown the operators
involving $R$ and $R_{\mu \nu}$ can be removed by field
redefinitions. These operators are redundant since the vacuum
equations of motion are $R_{\mu \nu}=0$. This means all physical
finite size operators can be built from the Weyl tensor $C_{\mu \nu
\rho \sigma}$ or alternatively from the Riemann tensor. Therefore,
the leading order finite size operators are composed of the electric
and magnetic components of the Weyl tensor squared
\cite{Goldberger:2005cd}. Hence, the worldline action including
these operators becomes
\begin{equation}
 S_{\textrm{BH}}[x,g_{\mu\nu}] = -m_0\int d\tau + \gamma_1 \int d\tau E_{\mu\nu} E^{\mu\nu} + \delta_1 \int d\tau B_{\mu_1 \ldots \mu_{d-2}} B^{\mu_1 \ldots \mu_{d-2}}, \label{eq:actionFSunch}
\end{equation}
where
\begin{equation}
E_{\mu \nu} = C_{\mu \alpha \nu \beta} \frac{dx^{\alpha}}{d\tau} \frac{dx^{\beta}}{d\tau} \ , \ \ \ \ \ \ B_{\mu_1 \ldots \mu_{d-2}} = \frac{1}{(d-2)!}\epsilon_{\alpha \mu_1 \ldots \mu_{d-3} \gamma \delta} {C^{\gamma \delta}}_{\beta \mu_{d-2}} \frac{dx^{\alpha}}{d\tau} \frac{dx^{\beta}}{d\tau}.
\end{equation}
Since our setup is static, the magnetic components of the Weyl
tensor do not contribute to any thermodynamic observable and so we
ignore the $\delta_1$ operator. If we consider a non-static problem
such as a scattering process, this operator may contribute since its
Wilson coefficient $\delta_1$ may be non-zero.

The $\gamma_1$ operator will give rise to new vertices coupling to
the worldline. The relevant worldline couplings will be $\phi\phi$
and $\phi\phi\bar \sigma_{ij}$. These are derived by expanding the
operator and the result is
\begin{equation} \label{eq:actionFSuncharged}
 S_{\textrm{BH}} \supset \gamma_1 (\partial_i \partial_j \phi)^2 - 2 \gamma_1 \bar \sigma_{ij} (\partial_i \partial_k \phi) (\partial_j \partial_k
 \phi),
\end{equation}
where we have neglected terms which vanish by the leading order
equations of motion. Terms with derivatives acting on $\bar
\sigma_{zz}$ in (\ref{eq:actionFSuncharged}) have also been omitted
since these do not contribute to our calculations.

To derive the thermodynamics including finite size effects, we need
to use a different approach compared to Section~\ref{sec:thermo}.
There we assumed the local temperature $T_0$ in (\ref{eqn:tempeqn})
is given by the temperature of a spherical uncompactified BH
(\ref{eq:T0rn}) and derived the asymptotic temperature using the
redshift. This methodology is justified when the BH is spherical.
However once finite size effects are allowed on the worldline, this
is no longer the case. Instead, we must first calculate the mass and
the tension of the system. With these relations we can derive the
remaining thermodynamic quantities, independent of the BH shape. The
redshift is also an observable so we compute it for completeness.

A direct computation of the tension from Feynman diagrams requires
we use the prescription of \cite{Chu:2006ce} and compute $\int
d^{d-1}\mathbf {x} \hspace*{1pt} T^{zz} = - \hat \tau L$. We will
however use the KK metric parametrization where the tension is
obtained from diagrams with one external $\bar \sigma_{zz}$. In this
case the sum of all diagrams yields $- \frac{1}{2} \hat \tau L
\bar\sigma_{zz}$.

\begin{figure}[tbp]
\centering
  \includegraphics{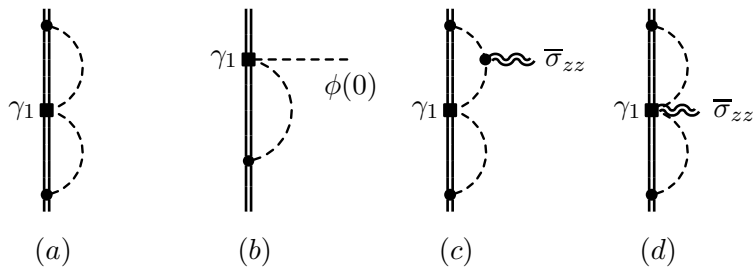}\\
  \caption{Feynman diagrams with the finite size operator $\gamma_1$. (a) Contribution to the ADM mass. (b) Redshift
  modification. (c) and (d) Tension contributions from $\gamma_1$.}
  \label{fig:nochargefinitesize}
\end{figure}

The Feynman diagrams needed to compute the contribution of the
$\gamma_1$ operator to the mass, the redshift, and the tension are
displayed in Fig.~\ref{fig:nochargefinitesize}. Their evaluation
requires some new sum integrals which are listed in
Appendix~\ref{Sec:IntegralsAppendix}. We obtain
\begin{align}
 \textrm{Fig.~\ref{fig:nochargefinitesize}$(a)$}&=  \tilde\gamma_1 m_0(d-1)(d-2)(d-3)^2 \mu^2, \\
 \textrm{Fig.~\ref{fig:nochargefinitesize}$(b)$}&=-2\tilde\gamma_1(d-1)(d-2)(d-3)^2 \mu^2,\\
 \textrm{Fig.~\ref{fig:nochargefinitesize}$(c)$}&=- \tilde\gamma_1 m_0(d-2)(d-3)^2 (d^2 - 4d + 5) \mu^2 \bar\sigma_{zz},\\
 \textrm{Fig.~\ref{fig:nochargefinitesize}$(d)$}&=-2\tilde\gamma_1 m_0(d-2)^2(d-3)^2\mu^2 \bar\sigma_{zz},
\end{align}
where we have defined $\gamma_1=\tilde{\gamma}_1m_0r_0^4$ so
$\tilde{\gamma}_1$ is dimensionless and the expansion parameter is
defined by
\begin{equation}
 \mu= \left(\frac{r_0}{L}\right)^{d-1} \zeta(d-1) \sim \lambda^{\frac{d-1}{d-3}}\, .
\end{equation}
The $\mathcal O(\mu^2)$ corrections to the mass, redshift and
tension are then
\begin{align}
 m & = m_0 \Big[1 - \tilde \gamma_1 (d-1)(d-2)(d-3)^2 \mu^2 \Big],  \\
 R & = 1 - 2 \tilde \gamma_1 (d-1)(d-2)(d-3)^2 \mu^2, \\
 \frac{\hat \tau L}{m_0} & = 2 \tilde \gamma_1 (d-1)^2 (d-2) (d-3)^2
 \mu^2,
\end{align}
where we did not include the $\mathcal O(\lambda)$ and $\mathcal
O(\lambda^2)$ contributions from Section~\ref{sec:thermo}. We see
that all corrections to thermodynamic quantities from the leading
finite size effects scale as $\mu^2 \sim
\lambda^{\frac{2(d-1)}{d-3}}$, where $\mu^2 \sim \lambda^4$ for
$d=5$, $\mu^2 \sim \lambda^3$ for $d=7$ and $\mu^2 \sim \lambda^p$
with  $2<p<3$ for $d > 7$.

From the Smarr relation and the first law we find the entropy
\begin{equation}
 S = S_0 \left[1 + 0\,\cdot \tilde \gamma_1 \mu^2
 \right],
\end{equation}
with $S_0$ given by (\ref{eqn:S0}) in the limit $\eta = 1$ and the
temperature
\begin{equation}
 T = T(L \rightarrow \infty) \left[1 -\tilde\gamma_1 (3d-5)(d-1)(d-2)(d-3)\mu^2\right],
\end{equation}
where $T(L \rightarrow \infty)$ is the uncompactified temperature of
(\ref{eq:T0rn}) with $\eta = 1$. We note that when finite size
effects are included, the entropy remains non-renormalized at
leading order. Moreover, we observe that the redshift $R$ does not
coincide with the renormalization of the temperature -- the local
temperature $T_0$ is not simply given by $T(L \rightarrow \infty)$.
In fact, we can now extract the local temperature including the
leading order finite size operator to be
\begin{equation}
 \tilde{T}_0=T/R=T(L\rightarrow\infty)\left[1-\tilde\gamma_1 (d+1)(d-1)(d-2)(d-3)\mu^2\right].
\end{equation}
The Wilson coefficient $\tilde \gamma_1$ is still undetermined and
its exact value must be extracted from a matching calculation at the
scale $r_0$. If possible, any matching calculation is best
undertaken in $d$ dimensions since there is no ambiguity when
extracting the Wilson coefficients. In $d$ dimensions the finite
size contributions to the thermodynamics have a different
dimensional scaling compared to the point particle contributions.
However, this is beyond the scope of the current work. Since $\tilde
\gamma_1$ is dimensionless, one would expect $\tilde
\gamma_1\sim\mathcal{O}(1)$; however, it can depend on the dimension
$d$.

\subsection{Including charge}

For a charged caged BH, we do not have a Ricci flat background, so
we need to consider the finite size operators with two derivatives.
In general, the effective BH action (\ref{eqn:BHaction}) including
all terms up to two derivatives is
 \bea
 S_{\textrm{BH}}[x_{\mu},g_{\mu\nu},a_{\mu}] &=&-m_0\int d\tau+Q\int dx^{\mu}a_{\mu}
 \non
 &+&\alpha_1 \int d\tau R + \alpha_2 \int d\tau R_{\mu\nu}\frac{dx^\mu}{d\tau} \frac{dx^\nu}{d\tau}
 \non
 &+& \alpha_3\int d\tau f_{\mu\nu}f^{\mu\nu}
 + \alpha_4\int d\tau f_{\mu\sigma} f^\sigma_\nu \frac{dx^\mu}{d\tau} \frac{dx^\nu}{d\tau} + \ldots,
 \label{finsizeSeff}
 \eea
where we have restored the time integrals for convenience and in the
last two lines we have introduced the lowest order operators
describing finite size effects. According to the power counting
rules \cite{Chu:2006ce, Kol:2007rx}, the Wilson coefficients of
these operators scale as $m_0 r_0^2$ if we power count $Q \sim m_0$.

Due to the presence of electromagnetism, operators involving the
Ricci tensor with coefficients $\alpha_1$ and $\alpha_2$ do not
vanish by the leading order equations of motion. However we can
still use the equations of motion, where $R\sim ff$ for the Ricci
tensor and scalar, see (\ref{eqn:scalar_curvature}). Thus we can
eliminate the $\alpha_1$ and $\alpha_2$ operators in favor of the
$\alpha_3$ and $\alpha_4$ operators. Alternatively, it is also
possible to redefine the coordinate system to eliminate $\alpha_3$
and $\alpha_4$, and the procedure is similar to the uncharged case
discussed in \cite{Goldberger:2004jt}.

We will eliminate the $\alpha_1$ and $\alpha_2$ operators and use a
different operator basis for the electromagnetic finite size
operators than in (\ref{finsizeSeff}) where the $\alpha_3$ and
$\alpha_4$ operators are expressed in terms of the electric
$e_{\mu}$ and the magnetic $b_{\mu_1...\mu_{d-3}}$ components of the
Maxwell field strength tensor
 \bea
 e_{\mu} &=& f_{\nu \mu} \frac{dx^\nu}{d\tau}~, \nonumber
 \\
 b_{\mu_1...\mu_{d-3}}&=&\frac{1}{(d-3)!}\frac{dx^\alpha}{d\tau}\epsilon_{\alpha\mu_1...\mu_{d-3}\beta\gamma}
 f^{\beta\gamma}~.
 \eea
For a static system the operator formed from the magnetic field will
not contribute to the thermodynamics, so we will neglect it. The
worldline action including the leading order static finite size
operator with two derivatives reads
 \bea
  S_{\textrm{BH}}[x_{\mu},g_{\mu\nu},a_{\mu}] &=&-m_0\int d\tau+Q\int dx^{\mu}a_{\mu} + \alpha \int d\tau e_\mu e^\mu  +
  \ldots\,.
  \label{eqn:op1}
 \eea

\begin{figure}[tbp]
\centering
  \includegraphics{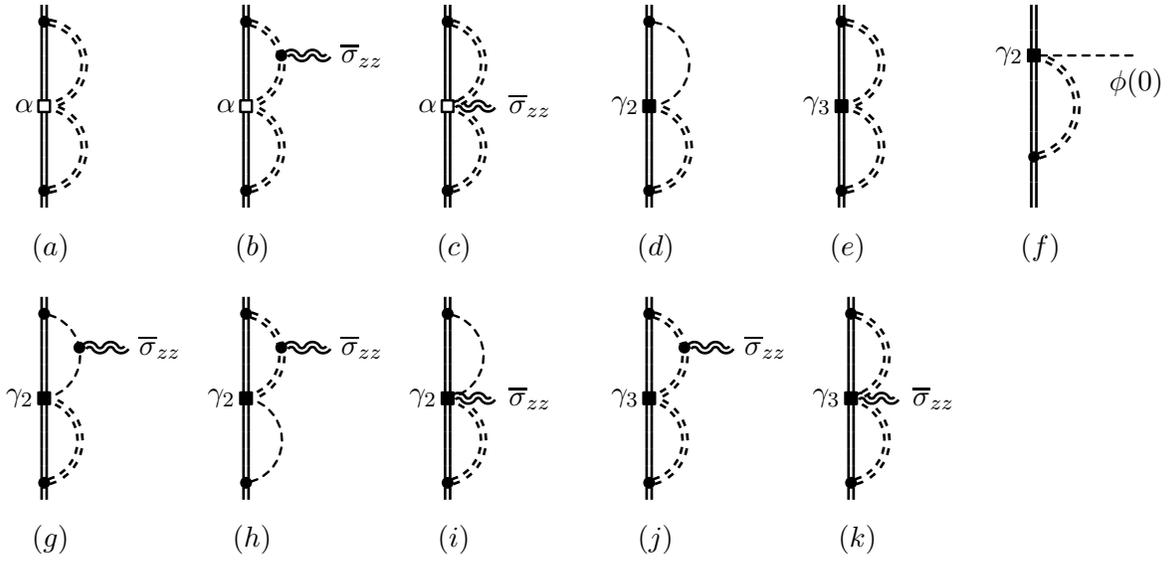}\\
  \caption{(a) Contribution to the ADM mass $m$
  from the finite size operator insertion. The open square denotes the vertex resulting from $\alpha \int d\tau e_\mu e^{\mu}$ insertion.
  (b) and (c) Tension contributions from $\alpha$ operator.
  (d) and (e) ADM mass renormalization diagrams resulting from the quadrupole finite size operators. The solid square vertex denotes the appropriate quadrupole finite size operators.
  (f) Contribution to the redshift of operator $\gamma_2$.
  (g)-(k) Tension contributions from the $\gamma_2$ and $\gamma_3$ finite size operators.
  }
  \label{fig:finitesize}
\end{figure}

We now compute the effect of the $\alpha$ operator. In the BH's rest
frame the $\alpha$ operator reduces to $- e^{2\phi/(d-3)}
\gamma^{ij} (\partial_i \varphi)(\partial_j \varphi)$. The Feynman
rule for the vertex coupling two $\varphi$'s to the worldline is
$2\alpha \textbf{k}\cdot \textbf{p}$, with both momenta incoming.
The corresponding mass renormalization diagram is shown in
Fig.~\ref{fig:finitesize}$(a)$, which scales as $\sim m_0
\lambda^{2(d-2)/(d-3)}$. Calculating this diagram shows that it
vanishes, and this has a clear interpretation. The finite size
operator in (\ref{eqn:op1}) encodes deformation due to the electric
dipole polarization of the BH. If we think of the problem as a line
of charged BHs at regular intervals in the covering space, as in
Fig.~\ref{fig:exactsol}, then the polarizing effect from a charge to
the right will be canceled by the equidistant charge on the left.
Therefore the overall polarizing effect is zero and there is no mass
renormalization from this worldline operator. Similarly, the
diagrams involving this operator and an external $\bar \sigma_{zz}$
which contribute to the tension, see Fig.~\ref{fig:finitesize}(b)
and (c), are found to vanish.

As a result, the first non-vanishing finite size contributions to
the BH thermodynamics come from operators with four derivatives,
which encode deformations due to induced quadrupole moments. As
explained above, in a static system we can form all relevant
operators using the electric field $e_\mu$, the electric components
of the Weyl tensor $E_{\mu \nu}$, and higher derivative
combinations. The static finite size terms in the action read at
order four derivatives
\begin{equation}
 S_{\textrm{BH}}[x, g_{\mu\nu}, a_\mu] = \gamma_1 \int d\tau E_{\mu\nu} E^{\mu\nu} - \gamma_2 \int d\tau (\nabla_\mu e_{\nu}) E^{\mu\nu} + \gamma_3 \int d\tau (\nabla_\mu e_{\nu}) (\nabla^\mu
 e^{\nu}),
 \label{quad_finsize}
\end{equation}
where the first operator coincides with the one in the uncharged
case of (\ref{eq:actionFSunch}).

The finite size terms in the action (\ref{quad_finsize}) lead to new
worldline vertices with two scalar fields, either gravitational or
electromagnetic. There are also vertices on the worldline with two
scalars and one $\bar \sigma_{zz}$ which contributes to the tension
calculation. The leading order correction to the ADM mass due to an
insertion of these operators introduced in (\ref{quad_finsize})
arises from the diagrams Fig.~\ref{fig:nochargefinitesize}$(a)$ and
Fig.~\ref{fig:finitesize}$(d)$ and $(e)$. Calculating the new
diagrams gives the renormalized mass
\begin{equation}\label{eqn:massfinitesize}
    m=m_0\left(1-\left[\tilde{\gamma}_1+\tilde{\gamma}_2 \frac{1}{2}
    \frac{Q}{m_0}\frac{d-2}{d-3}+\tilde{\gamma}_3\frac{1}{4}\left(\frac{Q}{m_0}\right)^2\left(\frac{d-2}{d-3}\right)^2\right](d-1)(d-2)(d-3)^2\mu^2\right),
\end{equation}
where we defined the dimensionless couplings $\tilde \gamma_i$ from $\gamma_i = \tilde \gamma_i m_0 r_0^4$.

For the redshift factor $R$ we compute
Fig.~\ref{fig:nochargefinitesize}$(b)$ and
Fig.~\ref{fig:finitesize}$(f)$ with the result
\begin{equation}\label{eqn:Rfinitesize}
    R=1-\left(2\tilde{\gamma}_1+\tilde{\gamma}_2 \frac{1}{2}
    \frac{Q}{m_0}\frac{d-2}{d-3}\right)(d-1)(d-2)(d-3)^2\mu^2,
\end{equation}
and the tension is computed via the diagrams
Fig.~\ref{fig:nochargefinitesize}$(c)$ and $(d)$ and
Fig.~\ref{fig:finitesize}$(g)$-$(k)$ to be
\begin{equation}\label{eqn:Taufinitesize}
    \frac{\hat \tau L}{m_0} = 2 \left[\tilde{\gamma}_1+\tilde{\gamma}_2 \frac{1}{2} \frac{Q}{m_0}\frac{d-2}{d-3}+\tilde{\gamma}_3\frac{1}{4}\left(\frac{Q}{m_0}\right)^2\left(\frac{d-2}{d-3}\right)^2\right](d-1)^2(d-2)(d-3)^2 \mu^2.
\end{equation}
In the charged case we have two dimensionful short distance
quantities, $m_0$ and $Q$, so in general the dimensionless couplings
$\tilde \gamma_i$ are functions of the order parameter $\eta$. Thus
the thermodynamics derivation of the remaining quantities $S$, $T$
and $\Phi$ becomes cumbersome and not insightful since we have to
keep derivatives of the couplings with respect to $Q$ and $m_0$.

In the extremal limit $\eta = 0$, we can use the exact solution of
Section~\ref{sec:exact_extremalsol} to undertake a matching
calculation for the undetermined Wilson coefficients. This is done
by comparing the exact results $m = m_0$, $\hat \tau = 0$, and $R =
(1 + \lambda)^{-1}$ to the expressions in the perturbative
expansions (\ref{eqn:massfinitesize}), (\ref{eqn:Rfinitesize}), and
(\ref{eqn:Taufinitesize}). The non-renormalization of the mass and
tension are seen to yield one relation between the Wilson
coefficients $\tilde \gamma_i$, since the linear combination of
coefficients in (\ref{eqn:massfinitesize}) and
(\ref{eqn:Taufinitesize}) is degenerate.

The remaining relation is derived from $R$. We note the $\mathcal
O(\mu^2)$ contribution must vanish, since it is non-analytic in
$\lambda$ in arbitrary dimensions. We find from
(\ref{eqn:Rfinitesize})
\begin{equation}\label{eqn:extremeRedReln1}
\left[\frac{\tilde{\gamma}_2}{\tilde{\gamma}_1}\right]_{\eta=0}=-
\sqrt{8 \frac{d-3}{d-2}} ~ \textrm{sign}(Q).
\end{equation}
Using this relation and $m=m_0$ or $\hat \tau = 0$ we get
\begin{equation}\label{eqn:extremeRedReln2}
\left[\frac{\tilde{\gamma}_3}{\tilde{\gamma}_1}\right]_{\eta=0}= 2 \frac{d-3}{d-2}.
\end{equation}
This constitutes a partial matching of the Wilson coefficients of
the finite size operators in the extremal limit.

The matching calculation presented above has several shortcomings.
The calculation of the thermodynamic properties of caged BHs cannot
yield the coefficient of any operator involving magnetic type
components, since these do not contribute for static systems.
Further, even in the case of the electric type operators, the
leading finite size operator with two derivatives $e_\mu e^\mu$ does
not contribute to the thermodynamics of caged BHs -- so we cannot
fix its Wilson coefficient $\alpha$. Lastly, while the three
electric type operators with four derivatives do contribute to the
thermodynamics, we are not able to fix all three Wilson coefficients
$\gamma_i$ uniquely, due to the degeneracy between the mass and
tension relationships.

We have not exploited the non-renormalization of the electrostatic
potential $\Phi$. We might be able to fix the Wilson coefficients
$\gamma_i$ uniquely if we could calculate the electrostatic
potential $\Phi$ in the EFT. If we succeeded in such a computation,
the $\mathcal O(\mu^2)$ contributions from the finite size operators
would most likely entail a linear combination of $\gamma_2$ and
$\gamma_3$. If this new equation was linearly independent of the
other equations derived here, this would imply $\gamma_1=\gamma_2
=\gamma_3 = 0$ in the extremal limit.

\section{Conclusion} \label{sec:concl}

Using an EFT approach we have analyzed the thermodynamic properties
of small compactified BHs carrying charge. We obtain the relevant
thermodynamic quantities to $\mathcal O(\lambda^2)$. Standard power
counting arguments show that up to $\mathcal O(\lambda^2)$, all
thermodynamic contributions arise from a point particle description
of the BH. This implies that to this order the horizon is spherical
and not deformed due to the presence of the compact dimension, and
the entropy is given by the area of a spherical horizon as for an
uncompactified charged BH.

In the extremal limit, we constructed an exact solution using the
standard methods of GR. From the exact solution, we find that the
mass $m$, tension $\tau$, and electrostatic potential $\Phi$ are
non-renormalized in the extremal limit, and the redshift is given by
a geometric series in $\lambda$. We use the extremal thermodynamics
and the uncharged perturbative results of \cite{Chu:2006ce,
Kol:2007rx} as checks of our perturbative results for the
thermodynamic properties and find agreement.

The leading finite size corrections to thermodynamic properties are
computed in both the charged and uncharged cases in terms of Wilson
coefficients of higher order operators. These coefficients need to
be determined by a matching calculation. In the uncharged case the
entropy does not acquire a correction due to the leading order
finite size operators. If the Wilson coefficient is non-zero, this
is interesting, since this means its area remains the same even
though the horizon is now deformed and no longer spherical.
Moreover, the local temperature and the temperature of an
uncompactified BH do not coincide when we include finite size
effects, which is a clear sign of a non-sphericity.

For charged caged BHs, we find the leading finite size operators
with two derivatives do not contribute to the thermodynamics. Also,
due to the presence of two short distance scales $Q$ and $m_0$, the
dimensionless Wilson coefficients can plausibly depend on their
ratio. We computed the leading finite size corrections to $m$, $R$,
and $\hat \tau$, which come from four derivative operators. In the
extremal limit, we undertake a matching calculation by comparing
with the exact solution and we constrain two out of the three Wilson
coefficients.

Our results contribute to the study of the phase diagram for black
objects with one compactified dimension. In particular, for
dimensions $d> 7$, the leading finite size operators yield the
dominant correction to the existing perturbative results at
$\mathcal O(\lambda^2)$. It would be interesting to perform a
complete matching of the Wilson coefficients of the finite size
operators. For the uncharged case, it may in fact be possible to
obtain an estimate for $\gamma_1$ from numerical data for $d>7$.

In the uncharged case, \cite{Chu:2006ce} performed an interesting
comparison between their perturbative results and the numerical
results of \cite{Wiseman:2002zc}. This comparison shows the
perturbative result at $\mathcal O(\lambda)$ matches the numerical
solution to 10-20\% all the way up to the point where the
BH/non-uniform BS phase transition occurs. Since we have shown the
finite size effects first enter at the same order as in the
uncharged case, we can expect our leading order results to behave
similarly. We can only make general inferences at this point, since
the details of a possible transition are not captured by the EFT.

Based on our result for the leading order tension
$\hat{\tau}\propto\eta^2\lambda$, for a given bare mass $m_0$ and
compactification $L$, the addition of charge to the caged BH will
reduce the tension. Due to the excellent agreement between the EFT
and numerical treatments of the uncharged caged BH, we speculate
that this property will also hold non-perturbatively. Therefore one
can reasonably expect the topology changing transition to a BS phase
to be delayed when charge is added to the BH.

Intuitively this increased stability of the BH phase can be
understood in the covering space. Here the transition to a BS will
occur when we increase the BH mass and the horizons of neighboring
image BHs overlap. If there was no horizon deformation this would
occur at values of the bare mass when the horizon radius is $r=L/2$.
However, the attraction between the neighboring image BHs can cause
this overlap to occur at lower bare mass due to horizon deformation.
Since like charges repel, the tension from the gravitational
attraction will be reduced by an electrostatic repulsion of the
image BHs -- making a charged caged BH relatively more stable than
an uncharged caged BH with the same bare mass.

So far there have been few studies of the BH/BS phase transition
with the inclusion of charge. We have covered the entire mass-charge
phase space for small $\lambda$ and our study provides a first
exploration of the BH side of the phase space. Due to the presence
of the additional parameter $Q$, a richer structure of the phase
diagram can be expected. Further study, either numerically or
analytically, towards completion of this phase diagram, is left for
future work. Investigation of the possible phases in the extremal or
near extremal limit may exhibit new phenomenology.

\subsection*{Acknowledgements}

We thank Walter Goldberger, Barak Kol, and Sourya Ray for useful
discussions. This work has been supported in part by grant
DE-FG-02-92ER40704 from the US Department of Energy (JBG \& AR) and
partially by The Israel Science Foundation grant (MS).

\appendix

\section{Table of Feynman integral sums} \label{Sec:IntegralsAppendix}

All integrals are derived using the techniques described in
\cite{Peskin:1995ev}, with the sums being Riemann zeta functions.
The first two results listed below, $I_0(L)$ and $I_1(L)$, were
derived in \cite{Chu:2006ce} and $I_2(L)$ was computed in
\cite{Kol:2007rx}. We list all integral sums used in this work below
\begin{align}
 I_0(L) & = \frac{1}{2L} \sum_{n=-\infty}^{\infty}\int\frac{d^{d-2}\textbf{p}_{\perp}}{(2\pi)^{d-2}}\frac{1}{\textbf{p}^2_{\perp}+(2\pi n/L)^2}
  =\frac{1}{16\pi G m_0}\frac{d-2}{d-3}\,\lambda,\label{eqn:sumintiden} \\
 I_1(L) & = \frac{1}{2L} \sum_{n=-\infty}^{\infty}\int\frac{d^{d-2}\textbf{p}_{\perp}}{(2\pi)^{d-2}} \left[\frac{2\pi n/L}{\textbf{p}^2_{\perp}+(2\pi n/L)^2}\right]^2
  = \left(2 - \frac{d}{2}\right) I_0(L), \label{eqn:I1} \\
 I_2(L) & = \frac{1}{2L} \sum_{n=-\infty}^{\infty}\int\frac{d^{d-2}\textbf{p}_{\perp}}{(2\pi)^{d-2}} \frac{\textbf{p}^2_{\perp}}{\textbf{p}^2_{\perp}+(2\pi n/L)^2}
  = \frac{(d-2)(d-3)}{L^2} \frac{\zeta(d-1)}{\zeta(d-3)} I_0(L), \label{eqn:I2} \\
 I_3(L) & = \frac{1}{2L} \sum_{n=-\infty}^{\infty}\int\frac{d^{d-2}\textbf{p}_{\perp}}{(2\pi)^{d-2}} \frac{(2\pi n/L)^2}{\textbf{p}^2_{\perp}+(2\pi n/L)^2}
  = - I_2(L) \label{eqn:I3}, \\
 I_4(L) & = \frac{1}{2L} \sum_{n=-\infty}^{\infty}\int\frac{d^{d-2}\textbf{p}_{\perp}}{(2\pi)^{d-2}} \frac{(2\pi n/L)^2 \, \textbf{p}^2_{\perp}}{\left[\textbf{p}^2_{\perp}+(2\pi n/L)^2\right]^2}
  = - \frac{1}{2}(d-2) I_2(L), \label{eqn:I4} \\
 I_5(L) & = \frac{1}{2L} \sum_{n=-\infty}^{\infty}\int\frac{d^{d-2}\textbf{p}_{\perp}}{(2\pi)^{d-2}} \frac{(2\pi n/L)^4} {\left[\textbf{p}^2_{\perp}+(2\pi n/L)^2\right]^2}
  = \frac{1}{2}(d-4) I_2(L), \label{eqn:I5}\\
 I_6(L) & = \frac{1}{2L} \sum_{n=-\infty}^{\infty}\int\frac{d^{d-2}\textbf{p}_{\perp}}{(2\pi)^{d-2}} \frac{(\textbf{p}^2_{\perp})^2} {\left[\textbf{p}^2_{\perp}+(2\pi n/L)^2\right]^2}
  = \frac{d}{2} \, I_2(L). \label{eqn:I6}
\end{align}

\section{Thermodynamics via the Helmholtz free energy}\label{Sec:Helmholtzappendix}

Here we demonstrate the calculation of the thermodynamics of
Section~\ref{subsec:Thermo} using the Helmholtz free energy $F$,
without using any assumptions about the entropy $S$ of the charged
caged BH. We start by defining the Helmholtz free energy
\begin{equation}\label{eqn:Helmhfree}
F=m-TS.
\end{equation}
The first law (\ref{eqn:1stlaw}) gives the differential relation
$dF=-SdT+\Phi dQ+\hat{\tau}dL$. We now derive the following
differential relations between the Helmholtz free energy and the
entropy $S$, the electromagnetic potential $\Phi$, and the tension
$\hat{\tau}$,
\begin{eqnarray}
\label{eqn:HelmS}   S &=& -\left(\frac{\partial F}{\partial T}\right)_{Q,L}=-\left(\frac{\partial F}{\partial m_0}\right)_{L,Q}\left[\left(\frac{\partial T}{\partial m_0}\right)_{L,Q}\right]^{-1}, \\
\label{eqn:HelmPhi}\Phi &=& +\left(\frac{\partial F}{\partial Q}\right)_{L,T}=-\frac{\partial(F,L,T)}{\partial(m_0,L,Q)}\left[\left(\frac{\partial T}{\partial m_0}\right)_{L,Q}\right]^{-1},\\
\label{eqn:Helmtau}\hat{\tau} &=& +\left(\frac{\partial F}{\partial
L}\right)_{T,Q}=
+\frac{\partial(F,Q,T)}{\partial(m_0,L,Q)}\left[\left(\frac{\partial
T}{\partial m_0}\right)_{L,Q}\right]^{-1}.
\end{eqnarray}
After the second equality in the above relations, we have rewritten
the thermodynamic relations to be functions of $m_0$, $L$, and $Q$,
since these are the bare parameters of the EFT.

At this stage we use the Smarr relation (\ref{eq:smarr}) to rewrite
$F$ eliminating the temperature-entropy term. This gives
\begin{equation}\label{eqn:HelmhSmarr}
    (d-2)F=m+(d-3)\Phi Q+\hat{\tau} L.
\end{equation}
This form is now amendable to solution, since we can form a
differential relation for $F$ through the above equations for $\Phi$
and $\tau$, (\ref{eqn:HelmPhi}) and (\ref{eqn:Helmtau})
respectively. Doing this allows us to rewrite (\ref{eqn:HelmhSmarr})
as follows
\begin{equation}\label{eqn:HelmSolve}
    (d-2)F=m+\left((3-d)\,Q\frac{\partial(F,L,T)}{\partial(m_0,L,Q)}
    +L\,\frac{\partial(F,Q,T)}{\partial(m_0,L,Q)}\right)\left[\left(\frac{\partial T}{\partial
    m_0}\right)_{L,Q}\right]^{-1}.
\end{equation}
We have now obtained what is obviously a first order multi-variable
partial differential equation for the Helmholtz free energy $F$. A
solution can be constructed for $F$ through iteration, first order
by order in $\lambda$ and then order by order in $\eta$ at a given
order in $\lambda$. After some calculation, we find $F$ to
$\mathcal{O}(\lambda^2)$, and this is given by,
\begin{equation}\label{eqn:HelmSoln}
    F=m_0\left[\left(1-\frac{d-3}{d-2}\,\eta\right)
    +\left(\frac{d-3}{d-2}\,\eta-\frac{1}{2}\,\eta^2\right)\lambda
    +\left(-\frac{d-3}{d-2}\,\eta+\frac{1}{2}\,\eta^2-\frac{1}{2}\frac{d-3}{d-2}\,\eta^3\right)\lambda^2\right].
\end{equation}
The entropy $S$, the electrostatic potential $\Phi$, and the tension
$\hat{\tau}$ are computed using the thermodynamic relations in
(\ref{eqn:HelmS})--(\ref{eqn:Helmtau}), and their results agree with
the ones stated in Section~\ref{subsec:Thermo}. Using a Legendre
transformation, the Gibbs free energy is  $G=F-\Phi Q$ and we find
agreement with (\ref{eqn:GibbsPotentia}).

\end{document}